\documentclass[prl,aps,showpacs,endfloats]{revtex4}
\usepackage{graphicx}
\begin{document}

\title{Probing pairing symmetry of $Sm_{1.85}Ce_{0.15}CuO_4$ via
highly-sensitive voltage measurements: Evidence for strong
impurity scattering}

\author{A. J. C. Lanfredi$^a$, S. Sergeenkov$^{a,b}$,
and F. M. Araujo-Moreira$^a$} \affiliation{$^a$Grupo de Materiais
e Dispositivos, Centro Multidisciplinar para o Desenvolvimento de
Materiais Ceramicos, Departamento de F\'{i}sica, Universidade
Federal de S\~{a}o Carlos, Caixa Postal 676 - 13565-905 S\~{a}o
Carlos, SP,
Brazil\\
$^b$Joint Institute for Nuclear Research, Bogoliubov Laboratory of
Theoretical Physics, 141980 Dubna, Moscow region, Russia}

\date{\today}

\begin{abstract}
Using a highly-sensitive home-made mutual-inductance technique,
temperature profiles of the magnetic penetration depth $\lambda
(T)$ in the optimally-doped $Sm_{1.85}Ce_{0.15}CuO_4$ thin films
have been extracted. The low-temperature behavior of $\lambda (T)$
is found to be best-fitted by linear $\Delta \lambda (T)/\lambda
(0)= \ln(2)k_BT/\Delta _0$ and quadratic $\Delta \lambda
(T)/\lambda (0)=\Gamma ^{-1/2}\Delta _0^{-3/2}T^2$ laws above and
below $T=0.22T_C$, respectively, which clearly indicates the
presence of d-wave pairing mechanism dominated by strong
paramagnetic scattering at the lowest temperatures. The best fits
produce $\Delta _0/k_BT_C=2.07$ and $\Gamma /T_C=0.25(T_C/\Delta
_0)^3$ for the estimates of the nodal gap parameter and impurity
scattering rate.
\end{abstract}


\maketitle

\section{Introduction}

 It is well established that most of the conventional low-$T_C$
 superconductors have s-wave pairing symmetry [1]. As for high-$T_C$ cuprates (HTC),
 the study of pairing symmetry in these materials still remains one of
 the most polemical and active fields of research.  A number of recent
 experiments, including phase-sensitive measurements [2,3], the angle
 resolved photoemission (ARPES) [4,5],  and  the Raman spectroscopy [6]
 have revealed that electron-doped HTC with nearly optimal
 doping have predominantly $d_{x^2-y^2}$ pairing symmetry. In particular, Kokales et al. [7],
   Prozorov et al. [8] and Snezhko et al. [9]  showed that the low
   temperature superfluid density of Ce-based magnetic superconductors
    $Pr_{2-x}Ce_xCuO_4$ (PCCO) and $Nd_{2-x}Ce_xCuO_4$ (NCCO) varies quadratically
    with temperature in the whole range of doping, in agreement with
    the theoretical prediction for a $d$-wave superconductor with impurity
    scattering. Moreover, recently  remeasured [10] magnetic-field dependence
    of the low-temperature specific heat of  optimally-doped ($x=0.15$) PCCO give
    further evidence in favor of $d$-wave-like pairing symmetry in this material
    at all temperatures below $4.5K$. We should also mention very
    interesting result [11] on anomalous change in the field dependence of the electronic
    specific heat in PCCO crystals from linear (at $T = 2K$) to nonlinear
    (at $T=3K$) temperature behavior which can provide plausible explanation for the previous
    conflicting experimental results on the pairing symmetry in the electron-doped cuprates.

    At the same time, much less is known about
    such electron-doped material as $Sm_{2-x}Ce_xCuO_4$ (SCCO). Since $Sm$ has a larger
    ion size than $Ce$, $Pr$ and $Nd$, it is expected that paramagnetic scattering
    contribution to low-temperature behavior of SCCO should be much stronger
    than in PCCO and NCCO. Indeed,  the penetration depth measurements
    on SCCO single crystals [12] have indicated that this magnetic  superconductor
    exhibits a rather strong enhancement of diamagnetic screening below $4K$ which
    (by analogy with PCCO and NCCO) could be responsible for a $d$-wave pairing
    scenario with rather strong impurity scattering.

To shed more light on the pairing symmetry of electron-doped
magnetic  superconductors, in this Letter we present a study on
the optimally-doped  $Sm_{1.85}Ce_{0.15}CuO_4$ (in the form of
thin films grown by the pulsed laser deposition technique) by
using a high-sensitivity homemade mutual-inductance bridge to
extract their penetration depths with high precision.

\section{Experimental Procedure and Extraction Method}

A few SCCO thin films ($d=200nm$ thick) grown by pulsed laser
deposition on standard $LaAlO_3$ substrates [7] were used in our
measurements. All samples showed similar and reproducible results.
The structural quality of the samples was verified through X-ray
diffraction and scanning electron microscopy together with energy
dispersive spectroscopy technique. To account for a possible
magnetic response from substrate, we measured several stand alone
pieces of the substrate. No tangible contribution due to magnetic
impurities was found. The critical temperature $T_C$ was
determined via the measured complex voltage output
$V_{AC}=V'+iV''$ as the temperature where $V'=0$ (see Fig.1).

The experimental bridge used in this work is based on the
mutual-inductance method. To measure samples in the shape of thin
films, the so-called {\it screening method} has been developed
[13]. It involves the use of primary and secondary coils, with
diameters smaller than the dimension of the sample. When these
coils are located near the surface of the film, the response
(i.e., the complex voltage output $V_{AC}$) does not depend on the
radius of the film or its properties near the edges. In the
reflection technique [14], an excitation (primary) coil coaxially
surrounds a pair of counter-wound (secondary) pick-up coils. If we
take the current in the primary coil as a reference, $V_{AC}$ can
be expressed via two orthogonal components, i.e.,  $V_{AC} = V_L +
iV_R$. The first one is the inductive component, $V_L$ (which is
in phase with the time-derivative of the reference current) and
the second one is the quadrature resistive component, $V_R$ (which
is in phase with the reference current). It can be easily
demonstrated that $V_L$ and $V_R$ are directly related to the
average magnetic moment and the energy losses of the sample,
respectively [15]. When there is no sample in the system, the net
output from the secondary coils is close to zero because the pick
up coils are identical in shape but are wound in opposite
directions. The sample is positioned as close as possible to the
set of coils, to maximize the induced signal in the pick-up coils.
An alternate current sufficient to create a magnetic field of
amplitude $h_{AC}$ and frequency $f$ is applied to the primary
coil by an alternating voltage source, $V_{in}$. The output
voltage of the secondary coils $V_{AC}$ is measured through the
usual lock-in technique [16].

To extract the profile of the penetration depth within the
discussed here method, one should resolve the following equation
relating the measured output voltage $V_{AC}$ to the $\lambda (T)$
sensitive sample features [14]:
\begin{equation}
V_{AC}=V'+iV''=i\omega I_p\int_0^{\infty}dx\frac{M(x)}{1+2x/Q}
\end{equation}
where $Q=i\omega G\mu _0(h_p+h_s)$. Here, $I_P$ and $\omega=2\pi
f$ are respectively the amplitude and the frequency of the current
in the primary coil, $h_P$ ($h_S$) is the distance from the
primary (secondary) coil to the sample, $G$ is the total
conductance of the sample, and $M(x)$ is a geometrical factor
[14]. Since the total impedance of the sample is given by [17]
$Z=R+i\omega L_k$, the expression for the sample's total
conductance reads:
\begin{equation}
G=\frac{1}{R+i\omega L_k}
\end{equation}
Here $L_k$ and $R$ are the kinetic inductance and resistance of
the sample, respectively. From the above equations it follows that
by measuring $V_{AC}(T)$ we can numerically reproduce the
temperature dependencies of both $L_k$ and $R$.

Finally, from the two-fluid model, the relation between $L_k$ and
$\lambda (T)$ for thin films (with film thickness $d\ll \lambda $)
is given by [1,17,18]:
\begin{equation}
L_k=\mu _0 \lambda \coth (\frac{d}{\lambda})\simeq \mu _0 \lambda
(\frac{\lambda}{d})
\end{equation}
It is worth mentioning that instead of the tabulation-based
procedure used before [19], in the present study we have
simultaneously determined $G(T)$ from Eq.(1) and extracted both
$L_k$ and $R$ using Eq.(2). Then from the temperature dependence
of $L_k$ we have obtained the seeking temperature dependence of
$\lambda (T)$.

\section{Results and Discussion}

 Fig.1 shows the typical results for the temperature dependence of the
 voltage output $V_{AC}(T)$ in a typical SCCO thin film with $T_C = 20.2K$.
 Fig.2 depicts the extracted
 variation of $\lambda ^2 (T)/\lambda ^2(0)$ for {\it all temperatures}
 obtained by using Eqs. (1)-(3).

Turning to the discussion of the obtained results, recall that for
conventional BCS-type superconductors with $s$-wave pairing
symmetry the superfluid fraction $x_s(T)=\lambda ^2(0)/\lambda ^2
(T)$  saturates exponentially as $T$ approaches zero. On the other
hand, for a superconductor with a line of nodes, $x_s(T)$ will
show a power-like behavior at low temperatures. More precisely,
for tetragonal symmetry (and neglecting dispersion in the $c$-axis
direction), the simple $d_{x^2-y^2}$ pairing state predicts a
linear dependence [20] $\Delta \lambda (T)/\lambda (0)= AT$ for
the low-temperature variation of in-plane penetration depth
$\Delta \lambda (T)=\lambda (T)-\lambda (0)$. Here
$A=\ln(2)k_B/\Delta _0$ with $\Delta _0$ being the amplitude of
the zero-temperature value of the $d$-wave gap parameter. At the
same time,  in the presence of strong enough impurity scattering
the linear $T$ dependence changes to a quadratic dependence
$\Delta \lambda (T)/\lambda (0)= BT^2$ where $B=\Gamma
^{-1/2}\Delta _0^{-3/2}$ with $\Gamma$ being the (unitary limit)
scattering rate, which is proportional to the impurity
concentration of the sample [21]. By trying many different
temperature dependencies (including both exponential and
power-like), we found that above and below $T=0.22T_C$ our
high-quality SCCO films are best-fitted by a linear (see Fig.3a)
and  quadratic (see Fig.3b) dependencies, respectively.  What is
important, the fits produce physically reasonable values of both
$d$-wave node gap parameter $\Delta _0/k_BT_C=2.07$ and
paramagnetic impurity scattering rate $\Gamma /T_C=0.25(T_C/\Delta
_0)^3$. Hence, our results confirm a universal pairing mechanism
in electron-doped magnetic superconductors with $d$-wave nodal
symmetry dominated by paramagnetic impurity scattering at the
lowest temperatures (for comparison, a strong-coupling BCS
behavior with $\Delta _0=2.07k_BT_C$ is shown in Fig.3 by dotted
line). It is also interesting to notice that boundary temperature
($T=0.22T_C$) which demarcates two scattering mechanisms (pure and
impure) lies very close to the temperature where strong
enhancement of diamagnetic screening in SCCO was observed [12]
attributed to spin-freezing of $Cu$ spins. Moreover, the above
crossover temperature remarkably correlates with the temperature
where an unexpected change in the field dependence of the
electronic specific heat in PCCO crystals was found [11]
attributed to the symmetry change from nodal to gapped. However,
to make a more definitive conclusion regarding this correlation
and the origin of the crossover, it is necessary to measure the
field-dependent contribution to electronic specific heat in our
SCCO films. And finally, comparing the above estimates of the node
gap parameter and impurity scattering rate deduced from our data
for SCCO with similar parameters reported for best PCCO crystals
[7] (with $2\Delta _0/k_BT_C=3.9$ and $\Gamma /T_C=0.13(T_C/\Delta
_0)^3$), we conclude that as expected larger $Sm$ ion indeed
produces larger contribution to impurity scattering.

\section{Summary}

By using a high-sensitivity home-made mutual-inductance technique
we extracted with high accuracy the temperature profiles of
penetration depth $\lambda (T)$ in high-quality optimally-doped
$Sm_{1.85}Ce_{0.15}CuO_4$ (SCCO) thin films. The low-temperature
fits of our data clearly demonstrated that this electron-doped
magnetic superconductor possesses a $d$-wave pairing symmetry.
More precisely, SCCO was found to follow a clean limit (with
linear temperature dependence of $\lambda (T)$) for $T>0.22T_C$
while dominated by strong paramagnetic impurity scattering (with
quadratic temperature dependence of $\lambda (T)$) for $T<
0.22T_C$.

We gratefully acknowledge financial support from Brazilian
agencies FAPESP and CNPq. We also thank S. Anlage, C. J. Lobb and
R. L. Greene from the Center for Superconductivity Research
(University of Maryland at College Park) for useful comments and
discussions.

\newpage

\begin{figure*}
\includegraphics[width=7.5cm]{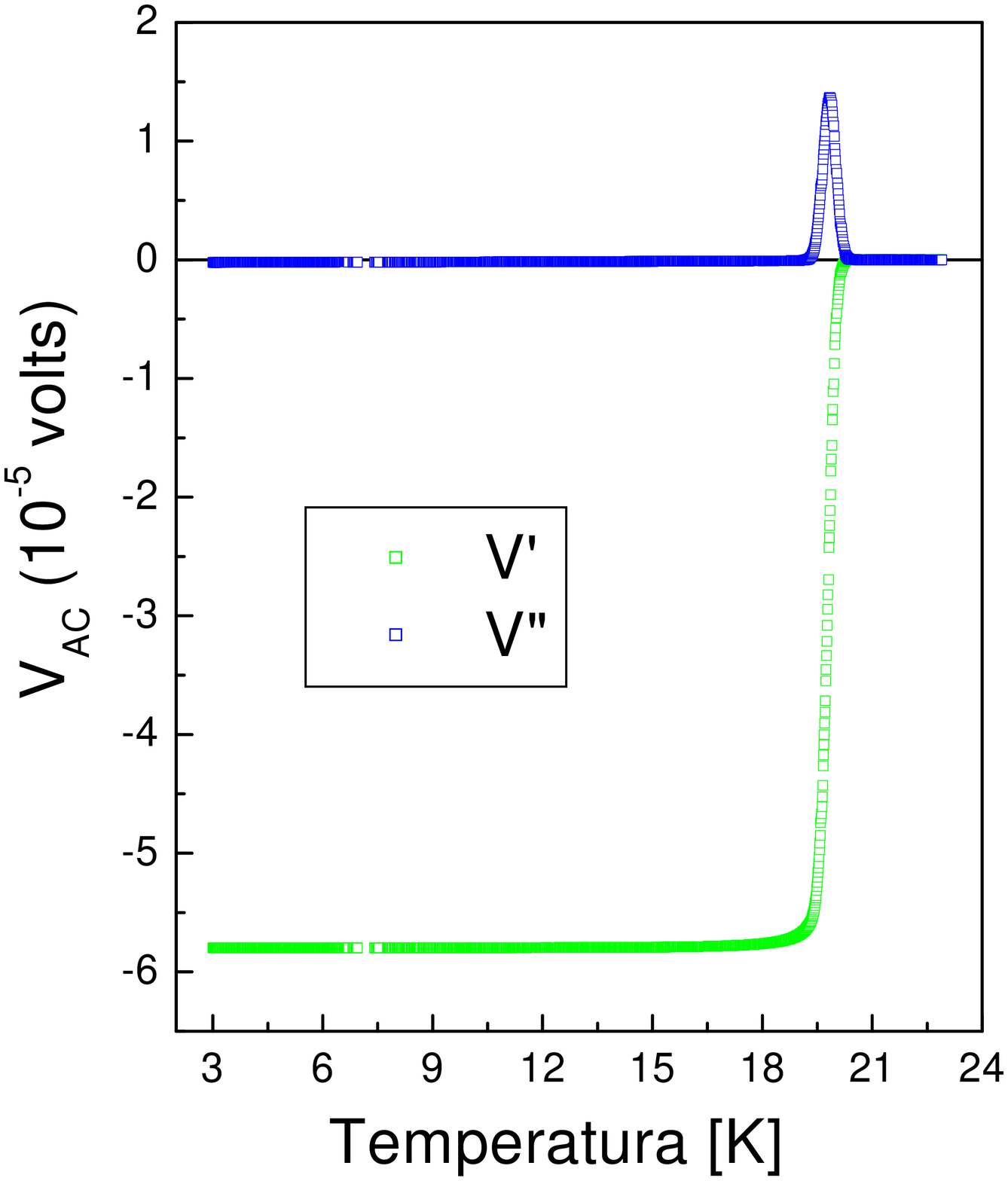} \vskip 0.5cm
\caption{Temperature behavior of the typical output voltages of
the secondary coils, $V_{AC}$, measured for a typical SCCO thin
film ($T_C =20.2 K$) under an alternate magnetic field of
amplitude $h_{AC}=100 mOe$ and frequency $f=55 kHz$.}
\end{figure*}

\begin{figure*}
\includegraphics[width=7.5cm]{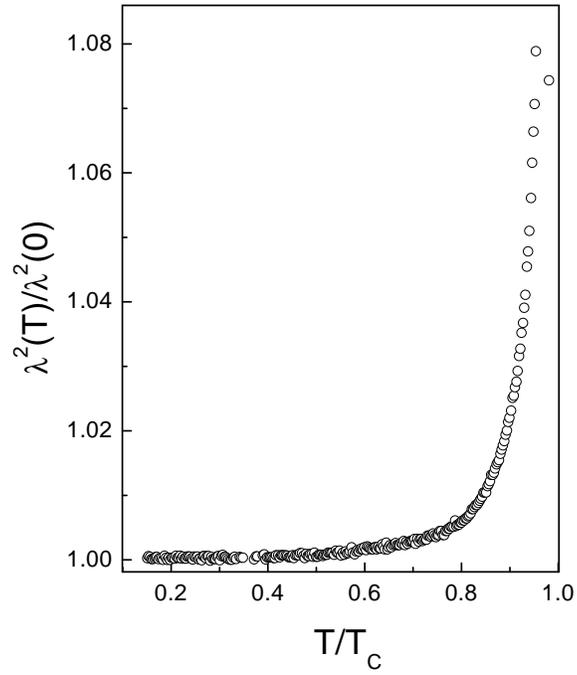} \vskip 0.5cm
\caption{Extracted from the output voltages $V_{AC}(T)$ variation
of $\lambda ^2(T)/\lambda ^2(0)$ for SCCO thin film as a function
of the reduced temperature using Eqs.(1)-(3).}
\end{figure*}

\begin{figure*}
\includegraphics[width=7.5cm]{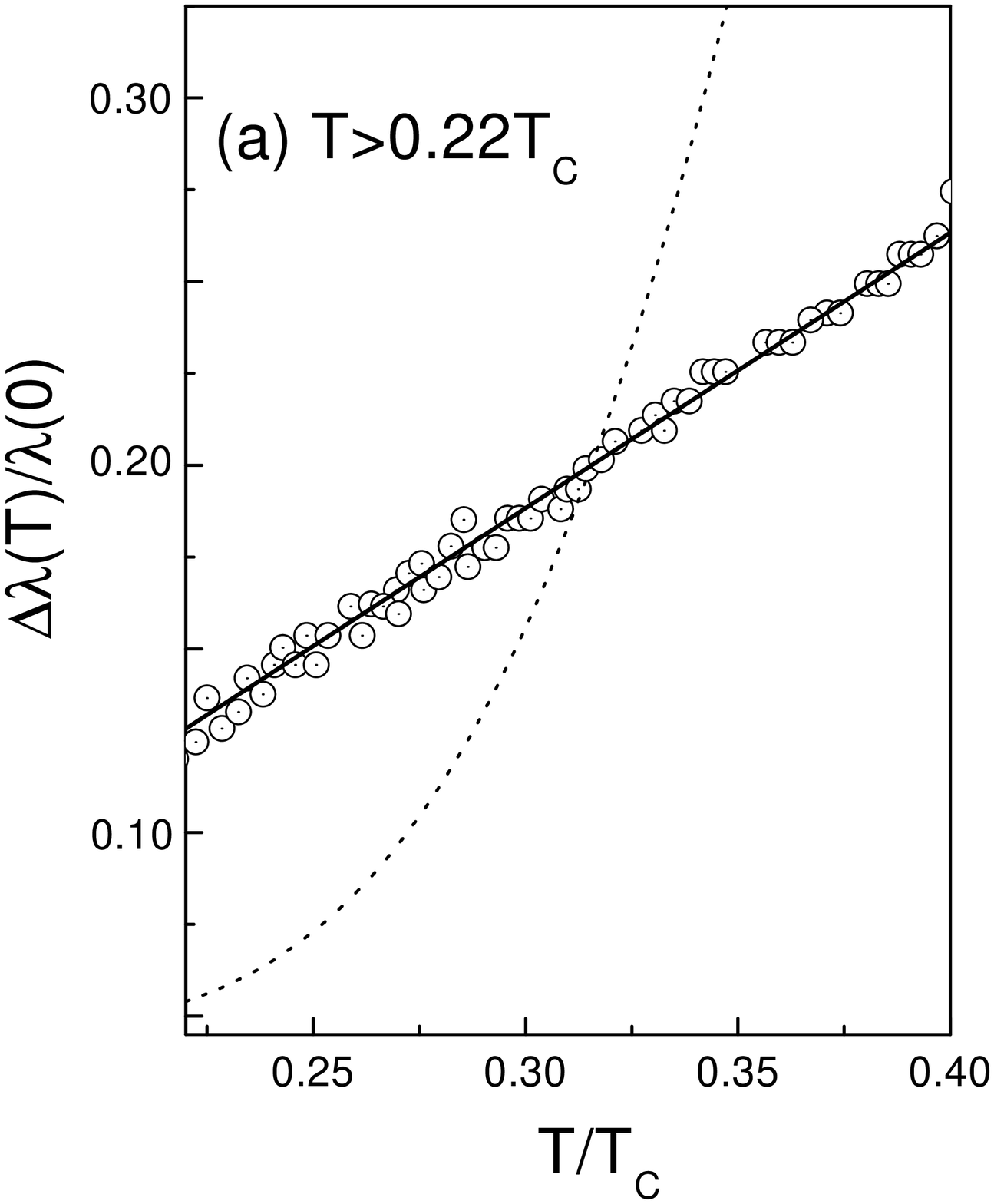}
\includegraphics[width=7.5cm]{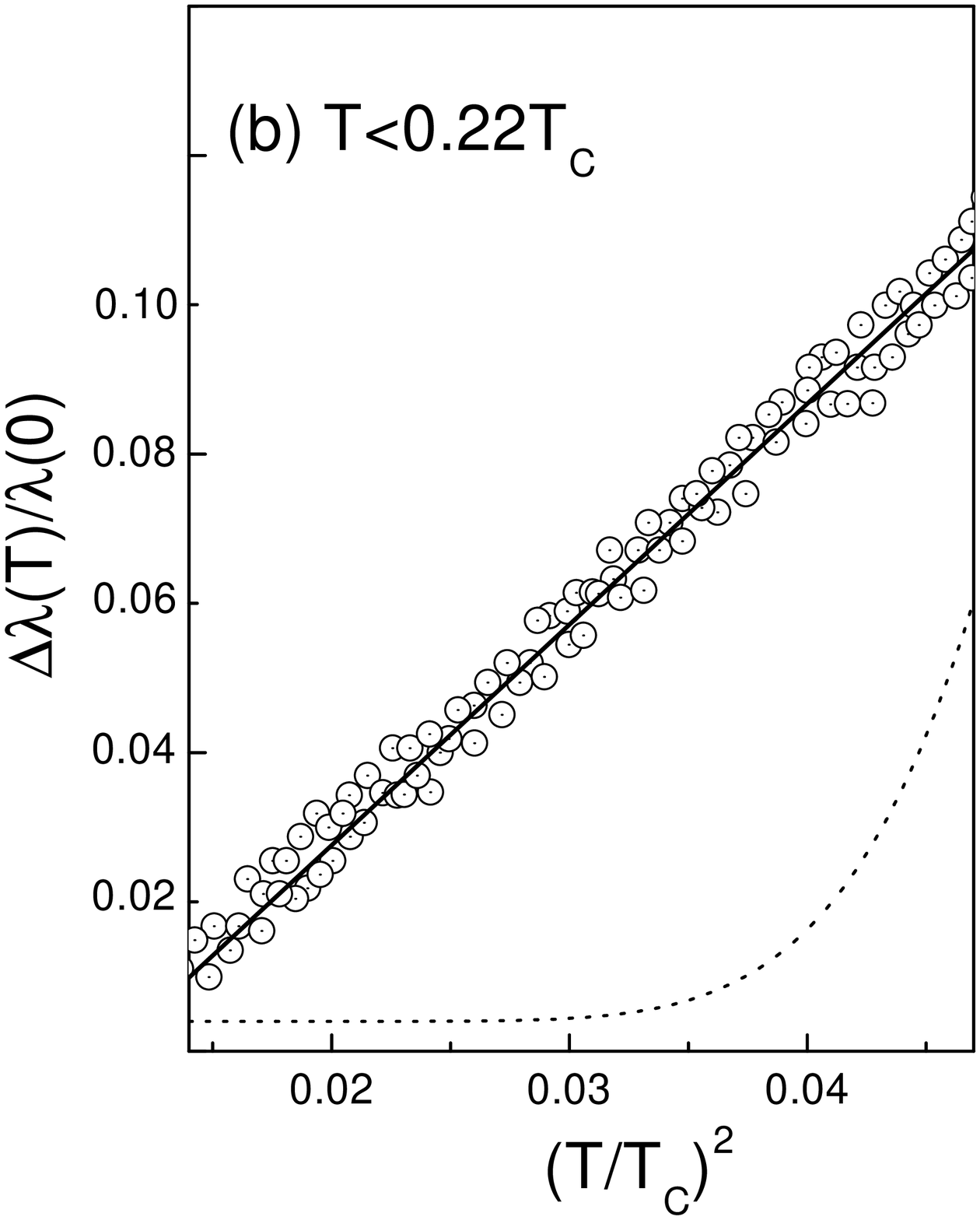} \vskip 0.5cm
\caption{Low temperature fits of the extracted variation of the
penetration depth $\Delta \lambda (T)/\lambda (0)$ in SCCO film
for two temperature regions: (a) $T>0.22T_C$ and (b) $T<0.22T_C$.
The dotted line shows a strong-coupling BCS behavior with $\Delta
_0=2.07k_BT_C$.}
\end{figure*}

\end{document}